  \documentclass[journal,twocolumn]{IEEEtran}

\usepackage[colorlinks,urlcolor=blue,linkcolor=blue,citecolor=blue]{hyperref}

\usepackage{amsmath}
\usepackage{amssymb}
\usepackage{graphicx}
\usepackage{url}
\usepackage{cite}
\usepackage{mathtools}
\usepackage{cuted}
\usepackage{etoolbox}
\usepackage{subcaption}
\usepackage{tikz}
 \usepackage{diagbox}
\usepackage{multirow}

\usepackage{enumitem}
 \newcolumntype{P}[1]{>{\centering\arraybackslash}p{#1}}

\usepackage{fourier} 
\usepackage{array}
\usepackage{makecell}
\usepackage{psfrag}

\usepackage{pifont}
\usepackage{soul}
\usepackage{soul,xcolor}

\begin{document}

	\title{Multi-band Wireless Networks: Architectures, Challenges, and Comparative Analysis}
	\author{Mohammad Amin Saeidi, {\em Graduate Student Member IEEE},  Hina Tabassum, {\em Senior Member IEEE,} and Mohamed-Slim Alouini, {\em Fellow IEEE} \thanks{M.A. Saeidi and H. Tabassum are with the Department of Electrical Engineering and Computer Science,  York University, e-mail: amin96a@yorku.ca, hinat@yorku.ca. M.-S. Alouini is with the Computer, Electrical, and Mathematical Sciences and Engineering Division, King Abdullah University of Science and Technology (KAUST), e-mail: slim.alouini@kaust.edu.sa. This research was supported by two Discovery Grants funded by the Natural Sciences and Engineering Research Council (NSERC) of Canada.}}
\raggedbottom

\maketitle	
\begin{abstract}
This paper presents the vision of multi-band communication networks (MBN) in 6G, where  optical  and TeraHertz (THz) transmissions will coexist with the conventional radio frequency (RF) spectrum. 
This paper will first pin-point the fundamental challenges in MBN architectures at the PHYsical (PHY) and Medium Access (MAC) layer,  such as unique channel propagation and estimation issues, user offloading and resource allocation, multi-band transceiver design and antenna systems, mobility and handoff management, backhauling, etc.  We then perform a quantitative performance assessment  of the two fundamental MBN architectures, i.e., {stand-alone MBN}  and {integrated MBN} considering critical factors like achievable rate, and capital/operational deployment cost.
{Our results show that stand-alone deployment is prone to higher capital and operational expenses for a predefined data rate requirement. Stand-alone deployment, however, offers flexibility and enables controlling the number of access points in different transmission bands.} In addition, we propose a molecular absorption-aware user offloading metric for MBNs and demonstrate its performance gains over conventional user offloading schemes. Finally, open research directions are presented. 
\end{abstract}

\begin{IEEEkeywords}
 6G, THz, multi-band, deployment cost efficiency, traffic offloading, hybrid transceivers
\end{IEEEkeywords} 	
	
\section{Introduction}
	To date, wireless technology has been developed primarily to broadcast data over radio frequency (RF) or sub-6GHz spectrum. However, the finite number of licensed RF bands cannot keep up with the future massive connectivity requirements. Extremely high frequencies (EHF) such as optical, millimeter-wave (mmWave), and Terahertz (THz) offer much wider transmission bandwidths with extreme data rates (in the order of multi-Gbps). Nevertheless, EHF transmissions are susceptible to unique channel propagation impediments (such as spreading loss, atmospheric loss, diffuse scattering, the impact of weather, and scintillation effects) resulting in smaller coverage zones and frequent switching among access points if a user is moving. 

In short, while RF can support broader coverage zones with minimal blockages and environment agnostic transmissions, EHF can support ultra-broad bandwidths resulting in ultra-high data rates, secure narrow beams, unlicensed low-cost spectrum, and compact transceivers due to small wavelengths and form-factor. Subsequently, there is no one-size-fits all spectrum solution, and all frequencies will naturally benefit from the co-existence in next generation wireless networks referred to  as \textit{multi-band} network (MBN). The MBN architecture is thus anticipated to be capable of opportunistically exploiting the best spectrum and adapts optimally for a variety of network applications and settings. A recent over-the-air multi-band demonstration showcased both sub-6 GHz and mmWave end-to-end 5G NR systems, using Ericsson’s 5G NR pre-commercial base stations (BSs) and Qualcomm Technologies’ 5G NR UE prototypes \cite{Prototype-Ericsson}. Also, Huawei developed SingleRAN Pro as a solution for 5G-oriented all-in-one sites, which can support sub-6 GHz and mmWave NR \cite{Huawei-MBN-Transceiver}.
 
\begin{figure*}
    \centering
    \includegraphics[scale=0.52, trim=4 4 4 4 ,clip]{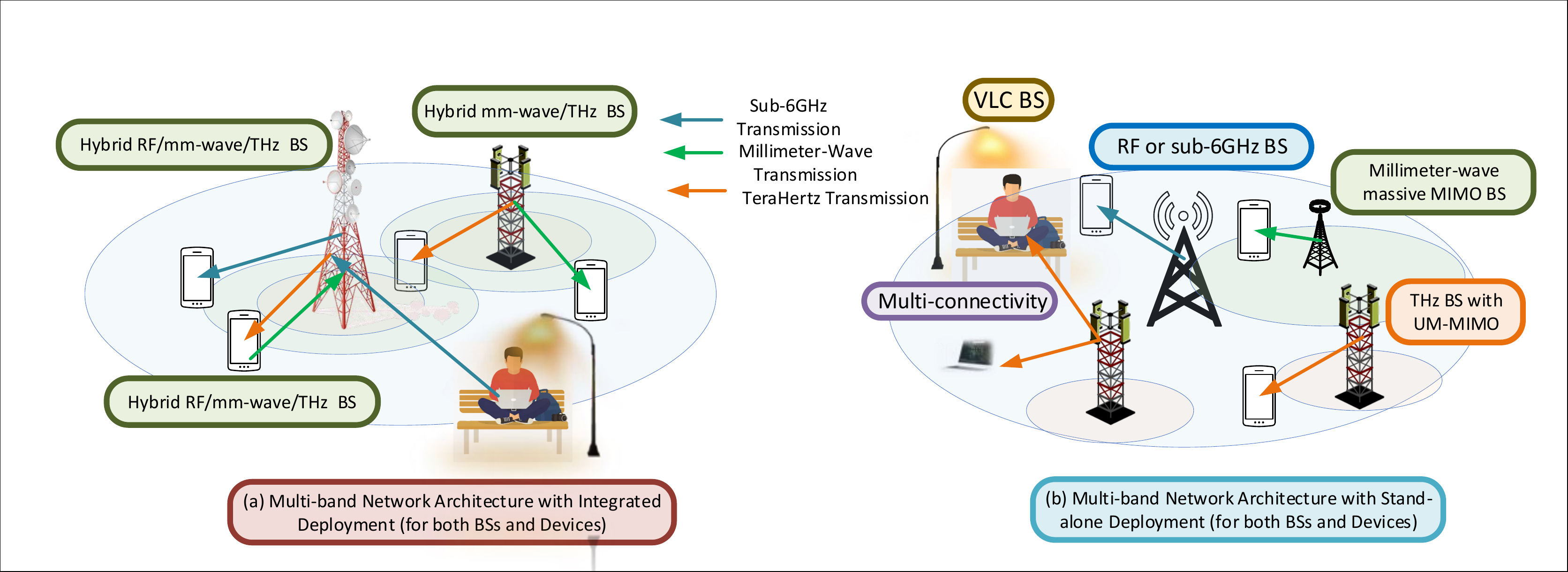}
    \caption{Vision of (a) {Int-MBN} architecture and (b) {SA-MBN} architecture in next generation wireless networks.}
    \label{fig:my_label}
\end{figure*}

 To this end, MBNs can be classified into two fundamental architectures, i.e.,
\begin{itemize}
    \item  \textit{Stand-alone MBN (SA-MBN):} The BSs with different frequencies are deployed separately and each BS can support only one spectrum at a time. This architecture offers the flexibility of deploying different types of BSs with varying densities to overcome propagation losses.
    \item \textit{Integrated MBN (Int-MBN):} Each BS can operate on more than one frequency band. Inspired by multi-homing users who are able to receive data at different frequency bands simultaneously \cite{multi-homing}, multi-band BSs in 6G can enable simultaneous transmission and reception at different frequency bands. 
\end{itemize}
The aforementioned MBNs can be realized with multi-homing user devices or stand-alone user devices as illustrated in Fig.~\ref{fig:my_label}. Another potential MBN architecture is a cascade configuration where RF and EHF links can be deployed on access and backhaul links, respectively. 

Some of the fundamental research along this line includes \cite{Coexisting-THz-TVT-2022,TanvirMobilityAware,MBN-RF-mmWave2}  considering EHF and RF bands jointly. {An energy-efficient solution for  RF/VLC systems is given in \cite{RF-VLC-Sylvester}.}
The authors in \cite{Transceiver3-THz-FSO} considered an integrated system with switching at the RF and free-space-optics (FSO) transceivers. {Note that conventional heterogeneous networks (HetNets) are always deployed in a stand-alone fashion. MBNs are different from HetNets as MBNs deal with diverse frequencies and channel propagation characteristics requiring innovation from the hardware design to  network PHYsical (PHY) and Medium ACcess (MAC) layer design.}
	

In this paper, we outline the key challenges in MBN architectures, including unique channel propagation and estimation issues, user offloading and resource allocation, multiband transceiver design and antenna systems, mobility and handoff management, and backhauling. Then, we compare the performance of  SA-MBN and Int-MBN in the presence of mobility and capital/operational deployment costs. Moreover, we propose a novel user offloading metric that adapts according to molecular absorption in the environment. Our findings indicate that SA-MBN can incur greater capital and operational expenses, despite allowing greater control over the number of access points in different transmission bands. It is also demonstrated that Int-MBN is more tolerant of user mobility than SA-MBN. Lastly, open research directions are discussed.

\section{Fundamental Challenges in MBNs}

In this section, we highlight the fundamental MBN challenges from the perspective of multi-band transceiver and antenna design as well as PHY  and MAC layer design issues.

\subsection{Multi-band Transceiver Design}
Designing multi-band transceivers is crucial in Int-MBNs. There are various methods that can be used for transceiver design, and each approach has its own set of challenges.
\subsubsection{Electronic up-conversion}
EHF signals are typically produced by using frequency multiplication in cascade and up-conversion techniques \cite{Transceiver1-CubSat}. 
Additionally, in a multi-band transceiver, frequency splitters can be applied to generate an intermediate frequency  as one of the desired output frequencies. However, operating at EHF bands can cause the CMOS technology components to exceed their optimal operating range, which  limits the output power.

\subsubsection{Optical down-conversion}
Multi-band transceivers can be realized by down-converting photonic signals to produce EHF signals. For instance, a Mach-Zehnder modulator generates signals at lower frequencies, while also maintaining the input photonic signals as the output. While the design of photonic-based transceivers is relatively simpler, they are sensitive to environmental conditions, such as temperature and humidity, which can affect their performance.

\subsubsection{Opportunistic frequency switching for Int-MBN}
A BS equipped with separate EHF and RF transceivers can employ both hard and soft switching to opportunistically exploit both links. Hard switching refers to the process of  switching frequencies using mechanical or electronic switches. This method is simple to implement, but lead to high power consumption and signal distortion caused by the switch transients. Soft switching, on the other hand, uses a combination of analog and digital signal processing techniques. This method can result in lower power consumption and less signal distortion, but it can also be more complex to implement and may have higher latency.

Hard switching  reduces the lifetime of the devices \cite{Transceiver3-THz-FSO}; whereas soft switching can outperform hard switching by providing lower outage probability. Note that multi-band transceivers and switching mechanisms are  critical in the Int-MBN deployment; whereas SA-MBN  can operate with  the users and BSs transmitting or receiving on a single predefined frequency.  The synchronization of time and frequency is also a challenge for successful decoding of messages transmitted on multi-band streams. 

\subsection{Multi-band Antenna Systems}
To enable high gain directional antennas, EHF allows to build dense arrays of antennas. There are several approaches to realize the arrays of tunable multi-band antennas as discussed below.

\subsubsection{Physically adjustable antennas}
based on micro or nano-electromechanical systems (MEMS/NEMS) can physically change the size of the radiating elements and achieve different resonant frequencies \cite{Transceiver1-6GSurvey}. The main challenge is to overcome the delay introduced by the control of the MEMS systems, particularly for fast data rates. 

\subsubsection{Electronically tunable antennas}
Another approach to building multi-band antenna systems is by using graphene-based 2D materials to construct software-defined nano-antenna arrays. The resonant frequency of these antennas can be adjusted by varying the graphene Fermi energy without any physical modifications. However,  graphene is primarily useful for the THz band, and other materials are needed to cover lower frequency bands. It is challenging to design a material that can operate at all frequencies.

\subsubsection{Distributed multi-band antenna systems}

To overcome the limited transmit power of existing EHF transceivers and antenna systems operating at EHF bands, one solution is to leverage  distributed antenna system. All BSs can cooperate to transmit the same data to a user with reduced transmission powers.  At the same time, the spatial diversity increases, leading to achieving higher data rates. However,  the communication overhead and control signals increases compared to centralized systems. Therefore, faster, scalable,  and robust algorithms are required to enable the use of cooperative distributed multi-band antenna systems.

\subsection{Channel Estimation and  Modeling}
Since the channel state information (CSI)  must be accessible at various nodes for resource management, faster, robust, and unique channel estimation solutions are critical that can capture the unique channel characteristics of each frequency band. Also, the channel coherence time (determined from the Doppler frequency shift)  varies significantly as a function of users' velocity and carrier frequencies. Thus, the channel estimation solutions should be faster to cope up with the reduce channel coherence time at EHF. In this context, deep learning can be applied for faster channel estimation and resource management as compared to traditional  computationally exhaustive solutions.  


The absence of precise channel models for EHF bands in various environments is also a fundamental challenge. The reason is that, when frequency increases, signals become more prone to atmospheric absorption losses caused by various gas molecules. Every band in the EHF spectrum has a transmission window (TW) where the molecular absorption is minimal. To accurately analyze and optimize the performance, unique channel models are required for different TWs \cite{TeraMIMO-ChannelModel}. Curve-fitting techniques can be used to create precise models of absorption loss for each TW. 

\subsection{Network Deployment and Planning}
Optimal network deployment and planning is critical due to additional degrees of freedom offered by various transmission frequencies.  For instance, increasing the number of RF BSs  increases interference; whereas, EHF needs more BSs to overcome propagation losses. Along another note, SA-MBN  enables deployment flexibility with higher capital, operational, and backhauling expenses, whereas Int-MBN requires sophisticated multi-band transceivers. {Based on the existing reports, one can postulate that the deployment and operational cost of hybrid BSs in Int-MBN is greater than that of SA-MBN \cite{Huawei-MBN-Transceiver}. {To be precise, using a BS with three sectors, each enabled with a separate frequency band, and utilizing a switching mechanism to exploit all three bands can be more expensive than a BS with just one sector that uses a single frequency band.} Thus, it is critical to investigate the trade-off between the deployment cost and performance of different MBN architectures.} 
    
\subsection{User Offloading and Mobility Management}
Since multiple frequency bands experience unique propagation characteristics and transmission bandwidths, user offloading methods need to adapt accordingly. Also, EHF  transmissions are prone to intermittent connectivity caused by narrower beams, beam misalignment, users' mobility, and small coverage zones, resulting in more handoffs and disconnections.
However, as EHF bands can deliver higher data rates, all of the required data can be transferred almost spontaneously as soon as a connection is established. In this context, MBNs are also crucial to   opportunistically exploit more stable RF/mmWave networks to maintain the desired QoS of moving users.   New traffic offloading mechanisms in MBNs are required that takes into account molecular absorption, beam misalignment and users' mobility. {In addition, the ability of a user to connect to two BSs in an SA-MBN deployment can result in additional overhead due to coordination between distant BSs.  This issue can be resolved by using Int-MBN where a user can associate with a single hybrid BS equipped with multi-band transceiver.}


\subsection{Resource Allocation and Backhauling}
    
{
Resource allocation, such as power and bandwidth, sub-channel assignment, user association, etc., is a key challenge in MBNs due to a mix of narrow-band and wide-band transmissions and environment-specific as well as transceiver-specific considerations. For instance, the considerations of  illumination constraint in VLC, molecular absorption-aware spectrum selection and user-offloading in THz, beam misalignment in FSO and THz, and the interference consideration of THz transmissions on passive satellite sensing are required. Furthermore, beam-squint phenomena in EHF, which is the spreading of beams in various sub-channels over different physical directions, impacts system performance due to the EHF's large bandwidth.
Also, traditional cyclic-prefix orthogonal frequency division multiplexing (CP-OFDM) may not be suitable in MBN due to a low peak-to-average power ratio and high susceptibility to the Doppler effect in EHF transmissions. Finally, the wireless backhauls must be opportunistic, environment-aware, and cost-efficient. For instance, SA-MBN requires more backhaul links than Int-MBN, and Int-MBN requires more complex backhauling strategies to support hybrid backhaul transmissions. Therefore, the trade-offs between cost and performance need to be investigated.}

\section{User Offloading in RF/THz MBN: A Case Study}


We also propose a novel simplified metric to perform efficient molecular absorption-aware offloading in MBNs and demonstrate its superior performance compared to the standard solutions.

\begin{figure}
   \centering
\includegraphics[scale=0.59]{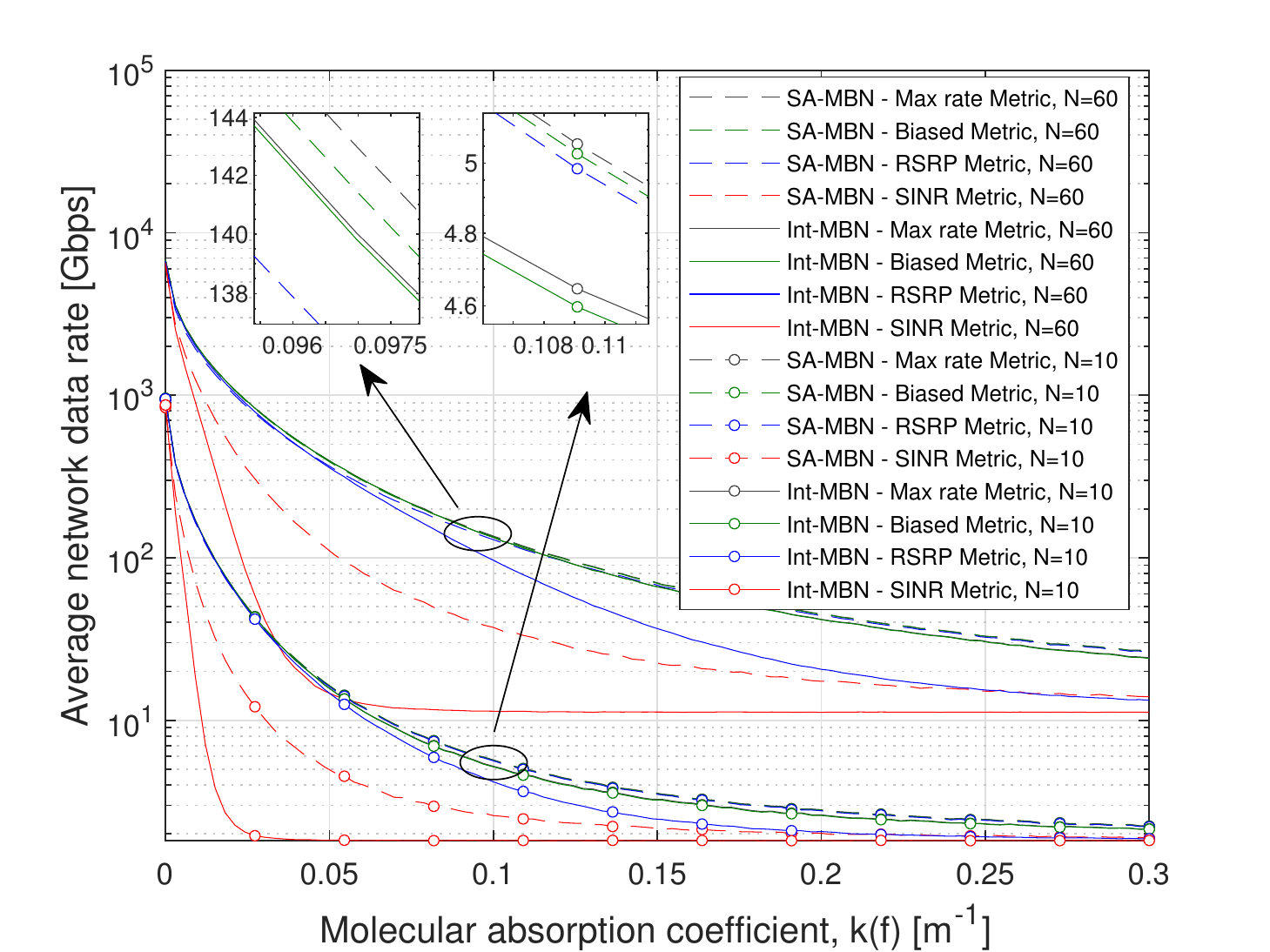}
\caption{{Network data rate versus molecular absorption coefficient for different number of BSs and UE-BS association metrics}}
\label{fig:RateVsNumberOfBSsDiffAssocMetrics_R2_Cmnt1}
\end{figure}


For SA-MBN deployment, we consider $N_T$ single antenna THz BSs (TBSs) and $N_R$ single antenna RF BSs (RBSs). For Int-MBN deployment, we consider $N_{\mathrm{Hyb}}$ hybrid BSs operating on both the THz and RF frequencies. We investigate the performance of a typical single antenna user. The RF and THz channels are modeled according to \cite{TanvirMobilityAware}.
The simulation parameters  are as follows. Antenna gain is 25 [dB] for THz transceivers and 0 [dB] for RF transceivers. The carrier frequency of THz is 1 [THz] and is 2 [GHz] for RF transmission. The probability of beam alignment with interfering TBSs is set to be 0.006. RF bandwidth is 40 [MHz], and the transmit power of RBSs and TBSs are 2 and 0.2 [Watts], respectively.

 {We propose a molecular absorption-aware bias for RBS and TBS as:
\begin{equation}\label{Bias}
{\text{Bias}}_{\mathrm{R}} = B_R  p_R^{Rx}, \ \ {\text{Bias}}_{\mathrm{T}} = B_T  p_T^{Rx}  \exp({K(f_T)d)},
\end{equation}
where $B_R$ and $B_T$ denote the RF and THz transmission channel bandwidths, respectively. Also, $p_R^{Rx}$, $p_T^{Rx}$, $\exp({K(f_T)})$, and $d$ denote the RF and THz received powers obtained using (1) and (3) in \cite{TanvirMobilityAware},}  molecular absorption coefficient at carrier frequency $f_T$, and the distance between BS and the typical user, respectively.

{Fig.~\ref{fig:RateVsNumberOfBSsDiffAssocMetrics_R2_Cmnt1} compares  various user offloading metrics in terms of network average data rate for both SA-MBN and Int-MBN. The maximum data rate metric based association outperforms others. However, the proposed bias-based association reduces CSI overhead and computational complexity, while attains a performance comparable to the maximum data rate for both SA-MBN and Int-MBN deployments.
The proposed scheme is followed by the reference signal received power (RSRP) metric and the SINR metric. The inferior performance of the SINR metric is due to not considering the impact of transmission bandwidths. Hence, when using SINR as the association metric, RF is selected most of the time, and even by  increasing the number of BSs, we cannot attain the high data rate provided by TBSs.}

 Considering different values of the molecular absorption coefficient, i.e., $K(f_T)$, reflects a variety of environments with varying humidity, temperature, and  weather. By increasing the molecular absorption, the Int-MBN performance degrades more sharply as user association probability to TBS is higher in Int-MBN.   By increasing the number of BSs from $N=10$ to $N=60$, the average network data rate improves due to the higher association to TBS in SA-MBN and THz mode of hybrid BSs in Int-MBN.

\section{SA-MBN vs Int-MBN: A Case Study}
This section provides a qualitative and quantitative comparison of the two MBN architectures in terms of BS deployment density, user offloading, spectral efficiency, and deployment cost efficiency.


\subsubsection{Deployment Density} 
\begin{figure}
   \centering
\includegraphics[scale=0.59]{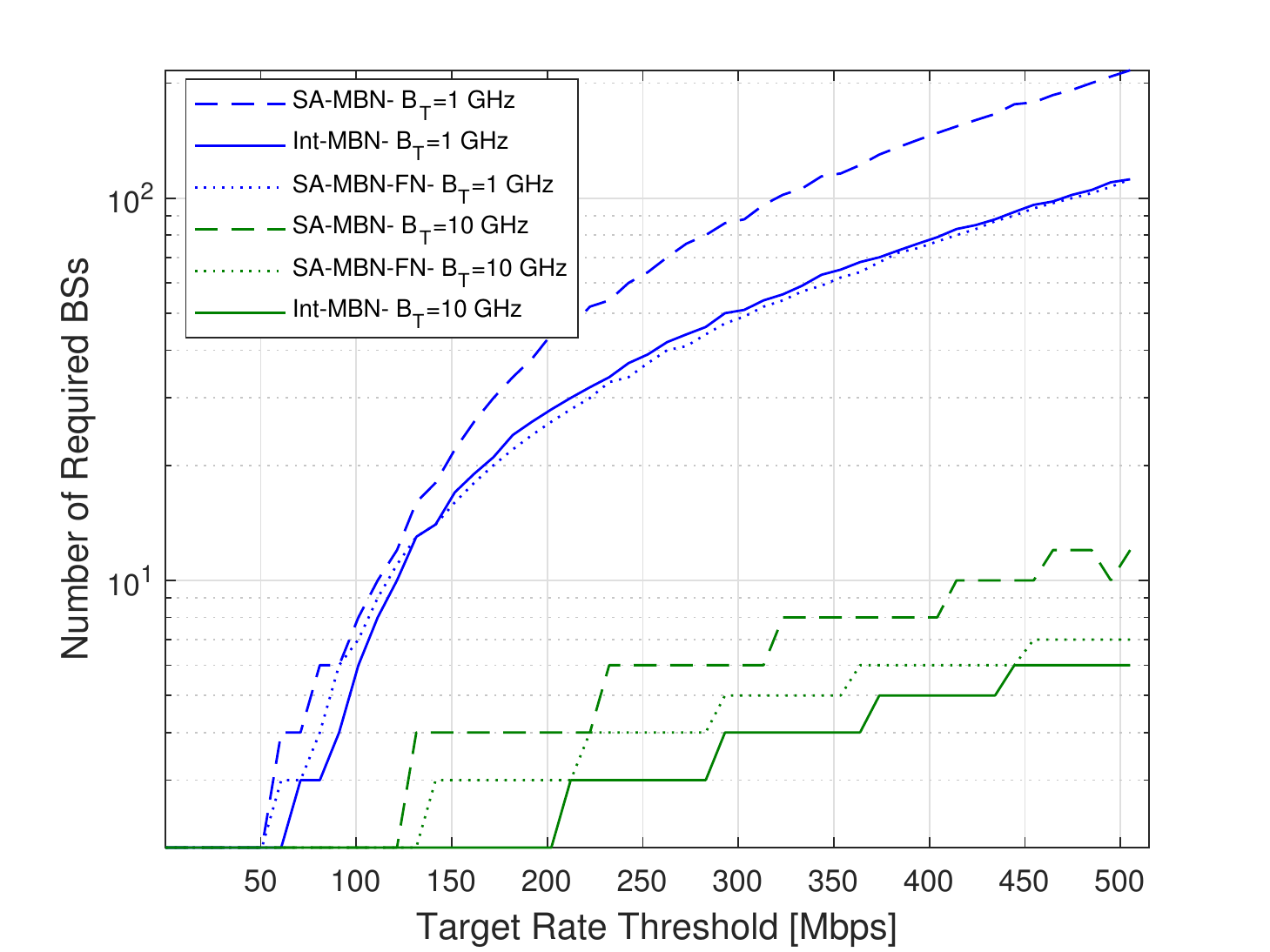}
\caption{The required number of BSs versus the target rate threshold. $R_{\max}=400 [m]$, $P_R^{tx} = 3 [\text{Watts}]$}
\label{fig:ReqBSsVsRateTh}
\end{figure}

Fig.~\ref{fig:ReqBSsVsRateTh} demonstrates the number of required BSs to reach the target rate requirement considering both SA-MBN and Int-MBN  for different THz transmission bandwidths, i.e., $B_T$. We first consider that the number of BSs in SA-MBN increases equally as Int-MBN, i.e., $N_T = N_R = N_{\mathrm{Hyb}}$. It can be seen that as we increase the target rate, SA-MBN requires a higher number of BSs compared to Int-MBN. Then, we investigate the case of SA-MBN with flexible number of BSs (\textit{SA-MBN-FN}), i.e.,
 selecting the number of RBSs and TBSs unequally. It can be seen that SA-MBN-FN can approach Int-MBN in terms of deployment density. That is, high interference in RF transmissions and low communication distance of THz transmissions can be compensated by raising the number of TBSs and lowering the RBSs (while minimizing the total number of BSs to a level similar to Int-MBN).  Nonetheless, the deployment cost can vary for the three configurations. 

Along another note, we observe that increased THz bandwidth improves the system performance, as it lets the typical randomly selected user achieves the required data rate with less number of BSs.

\subsubsection{Association to TBSs}
Fig.~\ref{fig:THz-Association} shows the  association probability of a user with TBSs as a function of the number of BSs. Note that RF association can simply be calculated by subtracting the association probability with TBSs from one. We note that a user in Int-MBN selects TBS with a higher probability than RBS. {The reason is that both frequency bands are available at each deployed hybrid BS. The user, thus, would select the one that provides a higher data rate, which is the THz. On the other hand, in SA-MBN, the user may not have access to closer TBS thus may associate to a nearby RBS.} The association of users with TBSs increases exponentially when the THz transmission bandwidth is large.

\subsubsection{Spectral Efficiency and Achievable Rate}

\begin{figure*}[t]
    \centering
     \subfloat[]{\includegraphics[scale=0.58]{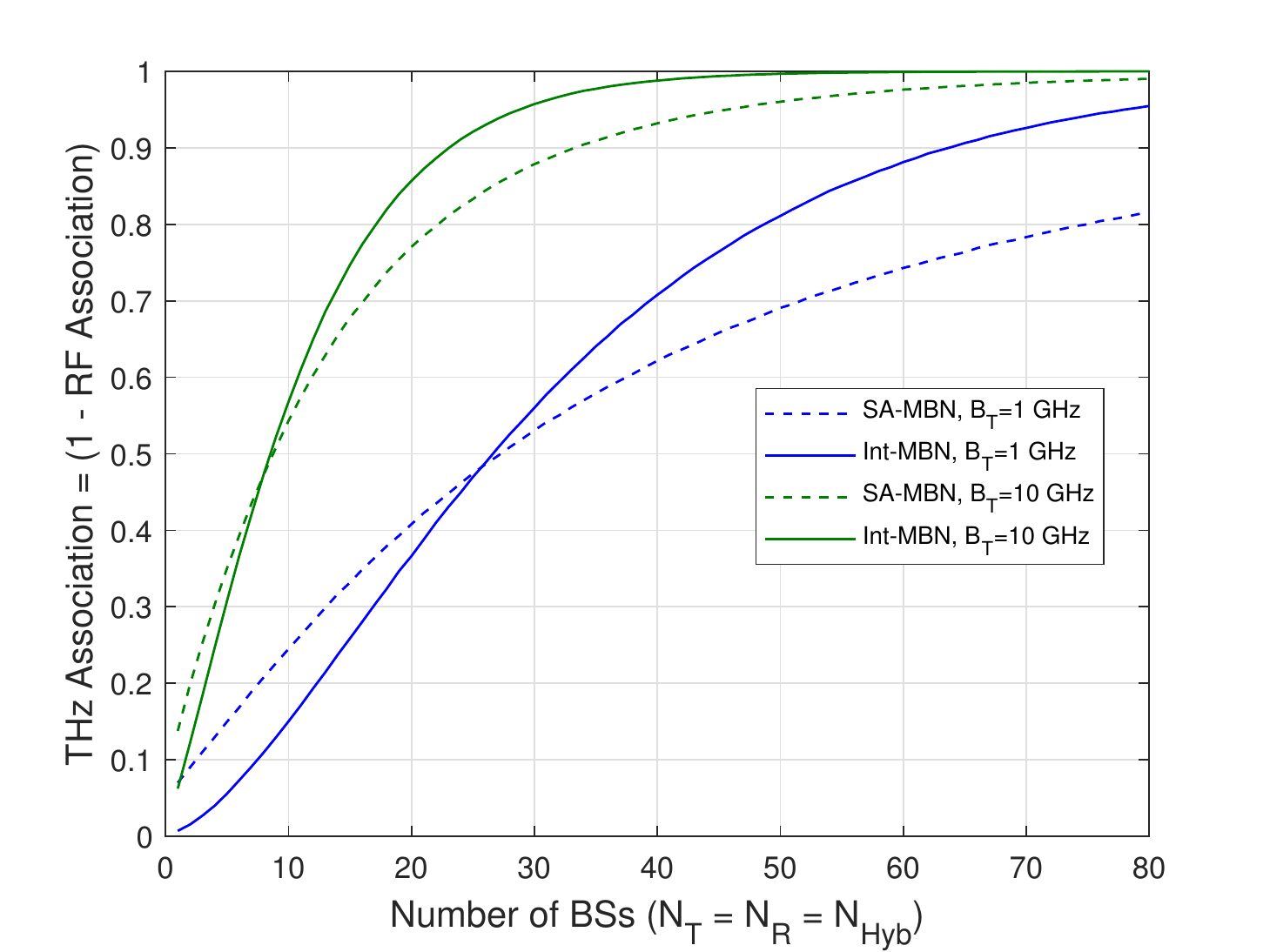}
    \label{fig:THz-Association}}
    \subfloat[]{\includegraphics[scale=0.58]{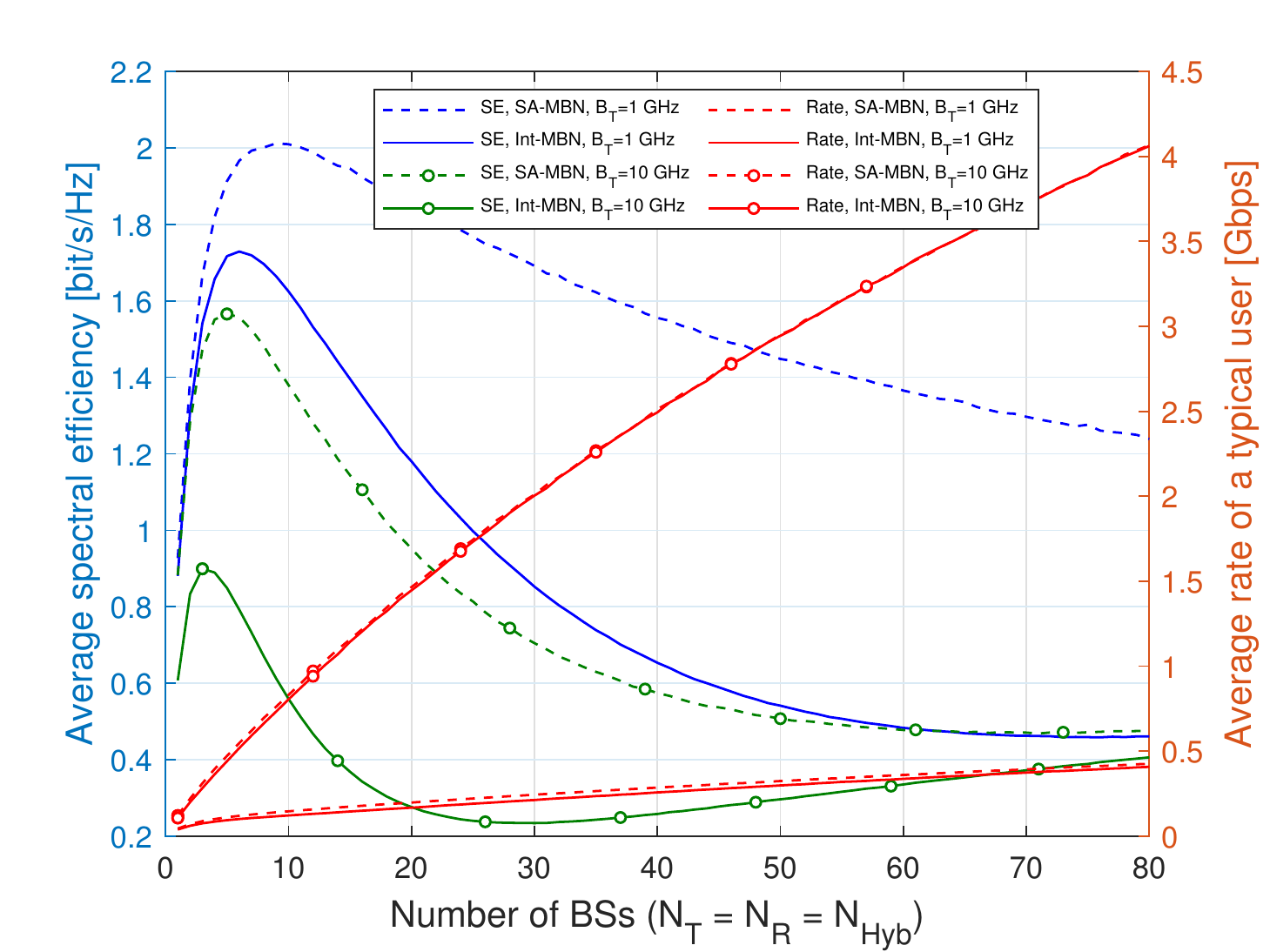}
    \label{fig:SEvsNumberOfBSs}}
    \vfill
    \hspace*{-0.2cm}
    \subfloat[]{\includegraphics[scale=0.58]{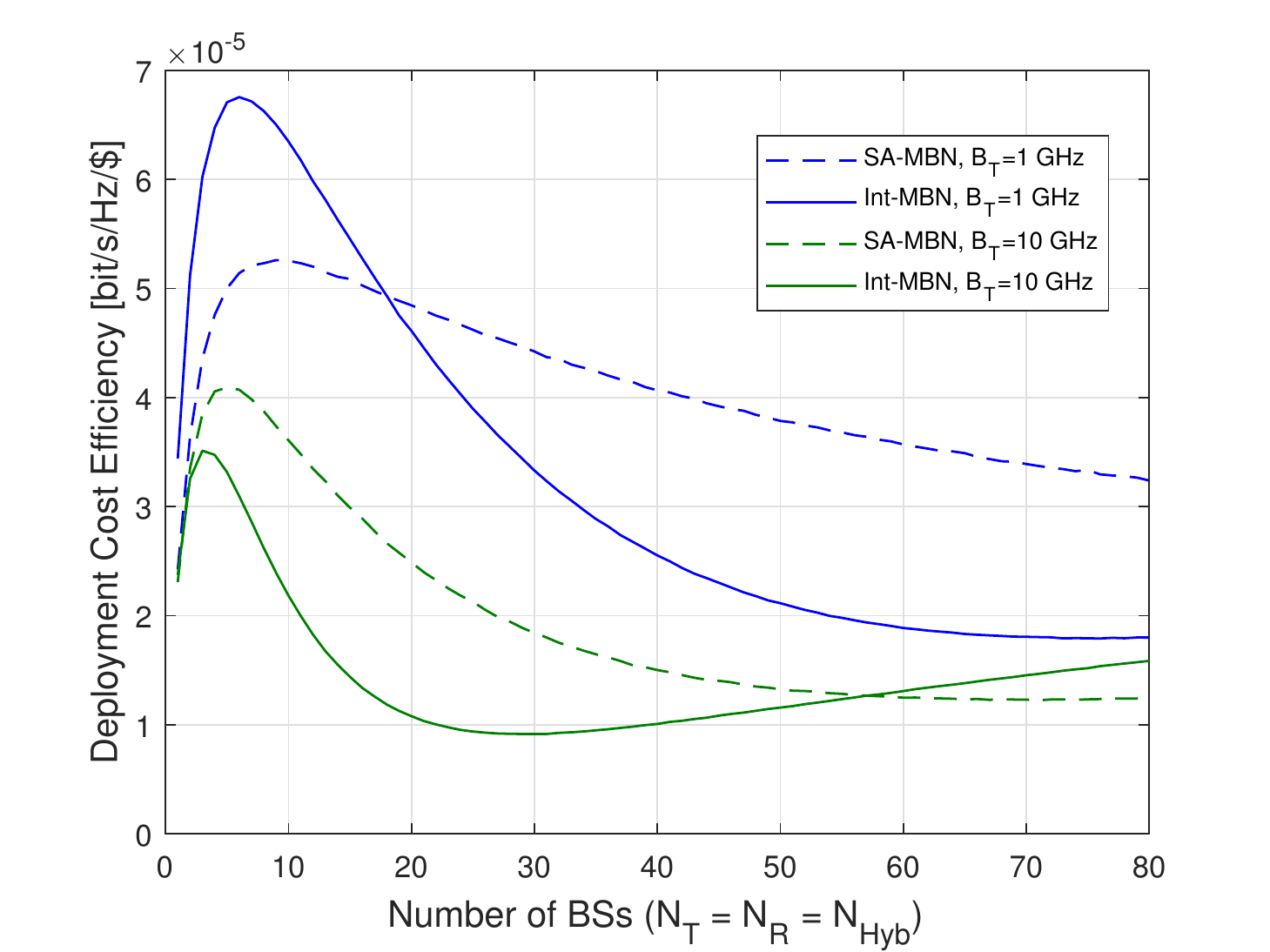}
    \label{fig:DeploymentEfficiencyVsBSsNumbers}}
    \hspace*{-0.7cm}
    \caption{{Impact of increasing the number of BSs at various THz bandwidth. (a) THz association probability, {(b) Spectral efficiency and data rate}, (c) Deployment cost efficiency. }}
    \label{fig:DCEfigures}
\end{figure*}

In Fig.~\ref{fig:SEvsNumberOfBSs}, the SE and data rate of each deployment for different THz bandwidths is demonstrated. 
The value of SE in the MBN is not expected to be huge as SE in THz band is typically smaller than that of RF because of the higher spreading losses and the existence of molecular absorption loss. However, the THz performance gains are achieved  mainly due to large THz bandwidth, which results in high achievable data rates. Subsequently, the larger the THz transmission bandwidth the higher is the achievable data rate (in the order of Gbps)  per user as is shown in Fig. 4(b). Also, we note that the SE  of SA-MBN is higher than that of Int-MBN. The reason is that in SA-MBN, BSs are more distributed in a given area. Therefore, the likelihood of finding a nearby BS is greater in SA-MBN.

In addition, we note that SE improves at first and then degrades as the number of BSs increases. When the number of BSs are less and more users are associating to RF BSs, the SE first increases due to increase in received signal power and then reduces due to increasing RF interference. This uni-modal shape is visible in the region where the association to RBS is higher than TBS.

\subsubsection{Deployment Cost Efficiency}

{
Deployment Cost Efficiency is defined as the ratio of achievable network spectral efficiency (SE) and the total investment in the installation of the network infrastructure, i.e., capital expenditure (CAPEX), which includes deployment of BSs, backhauling links plus the operational expenditure (OPEX), including maintenance, energy costs, and land lease.}

 We assume that TBSs have the same estimated deployment cost as mmWave BSs whose values are  available in \cite{GSMA-CAPEX,mason-CAPEX}. We estimate the CAPEX and OPEX as follows. The deployment of each  TBS needs \$38K, each RBS costs \$33K, and hybrid BS requires \$48K. Since hybrid BSs require sophisticated transceiver which operates at different frequencies, it is assumed that the transceiver costs more than the cost of BSs operating in a single frequency. For OPEX, we consider \$2.8K per RBS and \$2.7K per TBS and \$3.2K  per hybrid  BS. Note that these values are used as estimates for comparison and do not include all practical aspects. We define the deployment cost efficiency for SA-MBN and Int-MBN  as 
\begin{equation}\label{DCSA}
{\text{DCE}}_{\mathrm{SA}} = \frac{(N_R + N_T)\mathrm{SE^{SA}}}{N_R  C_R + N_T  C_T}, \ \  {\text{DCE}}_{\mathrm{Int}} = \frac{2 N_{\mathrm{Hyb}}\mathrm{SE^{Int}}} { N_{\mathrm{Hyb}} C_H}, 
\end{equation}
where $\mathrm{SE^{SA}}$ and $\mathrm{SE^{Int}}$ are the average user spectral efficiency of SA-MBN and Int-MBN, respectively. Also, $C_T$, $C_R$, and $C_H$ denote the cost of TBS, RBS, and hybrid BS, respectively.

Fig.~\ref{fig:DeploymentEfficiencyVsBSsNumbers} depicts  DCE  as the number of BSs increases. Since the cost of SA-MBN is higher than Int-MBN, scaling the SE of the two configurations with cost demonstrates the improved DCE of Int-MBN. Also, for $B_T = 10$ GHz, the THz association shows that the system rapidly tends to operate in THz, which imposes much less interference on the desired signals. Therefore, after 30 BSs, where THz is more than 0.95 percent used, without experiencing too much interference, DCE starts to increase again. Compared with SA-MBN, these results are primarily observable in Int-MBN since both THz and RF bands are available to be selected at each deployed BS.



\section{Future Trends and Prospects}
The performance of MBNs can be enhanced in different ways. Some potential directions are listed in the following:
\subsection{Machine Learning Solutions for MBNs}
Traditional optimization based resource allocation solutions are typically slow, computationally exhaustive, and not scalable. While this was acceptable in the previous decade, it is becoming practically difficult to adopt them in next generation large-scale MBNs. The reason is the short channel coherence time and a variety of unique channel impediments in EHF which necessitate faster, scalable, and robust resource management (i.e., power control, spectrum allocation, beamforming, beam tracking, antenna allocation, etc.) solutions.  In this context, deep learning solutions enable online resource management with deep neural networks that can be trained on environment specific CSI datasets. However, a fundamental challenge would be to incorporate convex and non-convex diverse QoS constraints and guarantee 100\% constraint satisfaction. Constraints associated with different transmission frequencies such as illumination constraint for VLC, molecular absorption-aware transmission windows in THz, and beam squint/misalignment in FSO and THz should be taken into account. A differentiable projection based framework is proposed very recently in \cite{alizadeh2023power}. 
{
It is critical to develop new deep learning techniques that are environment agnostic, i.e., the techniques where a given neural network can perform well for a different environment without significant retraining. In this context, meta-learning is a promising approach which enables a DNN to perform in a generalized environment with few-shot training.}


\subsection{Integrated Satellite-Aerial-Terrestrial MBNs}
EHF transmissions experience reduced propagation losses in space. For instance, THz transmissions do not incur significant molecular absorption effects at higher altitudes and free space, thus opening the door for longer communication ranges in space. Efficient aerial and space communications can mitigate the digital divide by connecting the unconnected in a seamless, safe, and resource efficient manner. It is thus crucial to investigate the opportunistic use of MBNs in such hybrid satellite-aerial-terrestrial networks. Since satellites/UAVs are mobile and experience larger propagation delays, MBN transmissions need to incorporate these features, along with the curvature of the atmosphere, pointing errors, atmospheric turbulence, the non-uniformity of atmospheric absorption losses, various antenna patterns, and the Doppler effects. 

\subsection{Reconfigurable Intelligent Surface (RIS)-aided MBNs}
The traditional RIS enables reconfigurable wireless environments by tuning reflection/refraction coefficients and phase shifts for a predefined transmission frequency. While RIS is beneficial in terms of reducing blockages and relaying the EHF transmission signals to farther distances, a given RIS will need to control the waves
at a variety of frequency channels in an MBN which is challenging. For example, the frequency and reflection phase of a graphene-based unit cell can be controlled via changes in its biasing voltage \cite{9920752}. Existing RISs with spatially splitted elements corresponding to each frequency transmission band is also of interest.


\subsection{Role of Distributed Antenna Techniques in MBNs}
In the considered Int-MBN and SA-MBN, massive antenna BSs can be deployed to further analyze the impact on handoff-aware network's capacity. The deployment costs, however, would be higher for both architectures, especially for SA-MBN, as it requires more BSs, compared to distributed MIMO systems (DMS). On the other hand, integration of MBNs with DMS offers more extensive coverage by leveraging EHF and sub-6GHz bands. In addition, Int-MBN can outperform when it comes to distributed systems owing to fewer, but more complex demands of backhauling links to the central coordinator of the system.

\subsection{Integrated Communication and Sensing in MBNs}
MBNs enable improved sensing and communication performance. 
Narrow beams in the EHF band can provide greater detection precision, whereas broad coverage in the RF band can detect distant targets. However, optimal resource allocation and beamforming design are required to effectively utilize the characteristics of each band.
In particular, hybrid BSs in Int-MBN can allocate different channels to either sensing or data transmission, while bi-static sensing can be implemented in SA-MBN using different BSs for a distinct purpose. 

\section{Conclusion}

In this paper, we comparatively analyzed two potential MBN architectures. We also highlighted the potential challenges of such configurations. In the simulation results, it shown that the proposed user offloading metric can achieve the performance of the maximum data rate association rule. 
It is shown that SA-MBN requires a higher number of BSs to achieve the same performance as Int-MBN for a given target data rate and equal proportion of RBSs and TBSs. 
However, SA-MBN can achieve the performance of Int-MBN if we allow unequal proportion of RBS and TBS deployments.

\bibliography{ref}
\bibliographystyle{IEEEtran}

\begin{IEEEbiography}
{Mohammad Amin Saeidi} received the M.Sc. degree in electrical engineering -- communication systems from the Amirkabir University of Technology, Tehran, Iran, in 2021. Currently, he is pursuing a Ph.D. degree in electrical engineering and computer science at York University. His research focuses on topics in wireless communications, terahertz communication, and intelligent reflecting surfaces.
\end{IEEEbiography}
\begin{IEEEbiography}
{Hina Tabassum} is currently a faculty member in the  Lassonde School of Engineering, York University, Canada. Prior to that, she was a postdoctoral research associate at University of Manitoba, Canada. She received her PhD degree from King Abdullah University of Science and Technology (KAUST) in 2013. She has been listed in the Stanford's list of the world's top 2\% researchers in 2021 and 2022 and has been recognized as N2Women: Rising Stars in Computer Networking and Communications in 2022. She is the founding chair of a special interest group on THz communications in IEEE Communications Society (ComSoc) - Radio Communications Committee (RCC). Her research interests include stochastic modeling and optimization of wireless networks including vehicular, aerial, and satellite networks, millimeter and terahertz communication networks.
\end{IEEEbiography}
\begin{IEEEbiography}
{Mohamed-Slim Alouini (Fellow IEEE)} was born in Tunis, Tunisia. He received the Ph.D. degree in Electrical Engineering from the California Institute of Technology (Caltech), Pasadena, CA, USA, in 1998. He served as a faculty member at the University of Minnesota, Minneapolis, MN, USA, then at the Texas A\&M University at Qatar, Education City, Doha, Qatar before joining King Abdullah University of Science and Technology (KAUST), Thuwal, Makkah Province, Saudi Arabia as a Professor of Electrical Engineering in 2009. His current research interests include the modeling, design, and performance analysis of wireless communication systems. 
\end{IEEEbiography}
\end{document}